\documentclass[twocolumn,amsmath,floatfix,prb,aps]{revtex4-1}

\usepackage{xcolor}
\usepackage{amsmath}
\usepackage{pifont}   
\usepackage{graphicx} 
\usepackage{dcolumn}  
\usepackage{bm}       
\usepackage{amsfonts} 
\usepackage{amssymb}  
\usepackage{multirow} 
\usepackage{romannum}
\usepackage{siunitx}
\usepackage{mathtools}

\usepackage{csquotes}
\MakeOuterQuote{"}

\usepackage{placeins}

\usepackage{braket} 

\usepackage{bbm}

    \renewcommand{\v}[1]{\bm{\mathrm{#1}}}

\begin{document}

\title{How closely does transient magnetic linear dichroism follow the spin moment?}
\author{E. I. Harris-Lee}
\affiliation{Max-Planck-Institut fur Mikrostrukturphysik Weinberg 2, D-06120 Halle, Germany}
\author{J.~K. Dewhurst}
\affiliation{Max-Planck-Institut fur Mikrostrukturphysik Weinberg 2, D-06120 Halle, Germany}
\author{P. Elliott}
\affiliation{Max-Born-Institute for Non-linear Optics and Short Pulse Spectroscopy, Max-Born Strasse 2A, 12489 Berlin, Germany}
\author{S. Shallcross}
\affiliation{Max-Born-Institute for Non-linear Optics and Short Pulse Spectroscopy, Max-Born Strasse 2A, 12489 Berlin, Germany}
\author{S. Sharma}
\email{sharma@mbi-berlin.de}
\affiliation{Max-Born-Institute for Non-linear Optics and Short Pulse Spectroscopy, Max-Born Strasse 2A, 12489 Berlin, Germany}
\affiliation{Institute for theoretical solid-state physics, Freie Universit\"at Berlin, Arnimallee 14, 14195 Berlin, Germany}

\date{\today}

\begin{abstract}

In highly out-of-equilibrium states of matter, such as those induced by a pump laser, the applicability of well established spectroscopic probes of magnetic order are called into question. 
Here we address the validity of x-ray absorption techniques in pump laser conditions, focusing on magnetic linear dichroism (MLD), a crucial probe of antiferromagnetic (AFM) order.
We directly compute the dynamics of the square of the spin moment and compare to those obtained via the MLD response. For AFM FePd the agreement between these distinct routes to the magnetic moment severely degrades at pulse fluences greater than $\sim1$~mJ/cm$^2$, indicating a breakdown of the MLD response as an accurate probe of the transient moment. This contrasts with the MLD for ferromagnetic FePt which reliably tracks the moment for fluences (and absorbed energies) up to an order of magnitude greater than the breakdown threshold for AFM FePd. The underlying microscopic reason for this we find to be increased laser induced excitations out of the $d$-band in AFM FePd, where this increase is made possible by the AFM pseudogap.
\end{abstract}

\maketitle

\emph{Introduction:}
Antiferromagnets exhibit a variety of desirable characteristics\cite{jungwirth_multiple_2018,olejnik_terahertz_2018,pal_setting_2022,jungwirth_antiferromagnetic_2016,verges_role_2020,fabiani_supermagnonic_2021,dannegger_ultrafast_2021,bossini_laser-driven_2019,nemec_antiferromagnetic_2018,wadley_electrical_2016,baltz_antiferromagnetic_2018}, for instance reversible current-driven switching at THz speed\cite{olejnik_terahertz_2018}, identifying them as key materials in future spintronics technologies.
The magnetisation dynamics of antiferromagnetic (AFM) materials has, furthermore, proved to be exceptionally interesting at ultrafast times\cite{fabiani_supermagnonic_2021,dannegger_ultrafast_2021,bossini_laser-driven_2019,nemec_antiferromagnetic_2018}, with pump laser induced switching of AFM to ferromagnetic (FM) order on femtosecond time scales being recently proposed \cite{dewhurst_laser-induced_2018}.
The early time spin dynamics of AFM matter thus presents rich possibilities for the control of magnetism by light.
Investigation of this ultrafast regime, however, relies on accurate measurement techniques to probe the transient magnetic order.
Unfortunately, the standard spectroscopic tool for probing ultrafast element-specific spin dynamics, magnetic circular dichroism (MCD), is inapplicable in situations of zero net magnetic moment. 
Magnetic linear dichroism (MLD), on the other hand, has quadratic dependence on the magnetisation, and so can be observed in both antiferromagnets and ferromagnets \cite{van_der_laan_experimental_1986,thole_strong_1985,Stohr2006}.
MLD therefore represents a key spectroscopic technique for investigating laser pumped AFM order, and indeed a number of recent experiments probing the time-evolution of AFM order via MLD have been performed\cite{thielemann-kuhn_ultrafast_2017,zhao_large_2021}.
It is thus timely to establish the reliability of this probe as a measure of {\it transient} moments in highly non-equilibrium AFM matter, and to identify what, if any, problems may arise in this context.

Both MCD and MLD are types of x-ray absorption spectroscopy, based on the principle that the magnetisation of a material can be estimated from its resonant x-ray absorption edges \cite{Stohr2006}.
For example, at the spin split $M$-edge or $L$-edge electrons are excited by a probe laser pulse from 3$p$ or 2$p$ states below the Fermi energy to empty $d$-states above the Fermi energy. 
The intensity of excitation allows one to count the number of empty $d$-states and, given that the $d$-band can always accommodate 5 spin up and 5 spin down electrons, this allows one to determine the moment on an atom.
For a circularly polarized pulse the dichroic response (the difference in response to right and left circularly polarized light, i.e the MCD) is proportional to the magnetic moment, while for a linearly polarized probe pulse the difference between the parallel and perpendicular response functions, relative to the quantization axis, is proportional to the square of the magnetic moment.
One should stress that this estimation of the moment from empty state counting, while exact for an isolated atom, becomes approximate in solids for which the lower symmetry crystallographic environment and concomitant hybridization of angular momentum channels complicates the "counting principle" underlying MCD\cite{Stohr2006,ebert_l-edge_1996}. Nevertheless, numerous experiments attest to the success of MCD and MLD as probes of magnetic order in the ground state.

An investigation of the measurement of transient MCD \cite{sharma_computational_2022} found that pulses with higher incident power densities trigger significant failures of the MCD to track the transient magnetization in the elemental ferromagnets Co and Ni. 
This failure could be attributed to excitation of charge from bands of $d$-character to bands of delocalized character, made possible by the hybridization of $d$ with relatively delocalized states in solids\cite{sharma_computational_2022}, resulting in violations of transient $d$-band state counting and overestimation of demagnetization by MCD. It is thus crucial to ask how this physics manifests in the MLD response of antiferromagnets, in particular given the quite different electronic structure in transition metal antiferromagnets.

To study the ability of MLD to accurately track the transient spin moment, we compare a prototypical FM, FePt, to its closest AFM analogue, FePd. Strikingly, we find that the failure of MLD to track the moment in AFM FePd occurs at an order of magnitude less absorbed energy than required to induced failure of the MCD response in FM FePt. The underlying microscopic reason we identify to be dramatically increased loss of $d$-band moment to delocalized states at the same absorbed energy, driven by increased $d$-$sp$ hybridization at the AFM pseudogap. As a corollary, we confirm that both MCD and MLD remain reliable tools to track the $d$-band moment on these timescales, even when failing to track the total moment.

\emph{Methods}: 
In the present work we use the fully \textit{ab-initio} method of time-dependent density functional theory (TD-DFT).
This method has been shown to be very successful at describing the dynamics of spins in laser pumped solids \cite{willems_optical_2020,dewhurst2020,hofherr2020}.
Within TD-DFT and its linear response extension \cite{RG1984,my-book,sharma2011,sharma_computational_2022} two distinct schemes can be used to obtain the magnetic moment.
In Scheme (i) the magnetization density, $\mathbf{m}(\mathbf{r},t)$, can be calculated directly from the expectation of the spin operator, $\hat{\mathbf{s}}$,
\begin{equation}
\mathbf{m}(\mathbf{r},t) = \sum_j \psi_j^*(\mathbf{r},t) \ \hat{\mathbf{s}} \ \psi_j(\mathbf{r},t),
\label{eqn:expectation}
\end{equation}
where $\psi_j$ is the spinor-valued wavefunction with index $j$ obtained from solution of the TD-DFT Hamiltonian \cite{my-book}. 
The magnetic moment, {\bf M}(t), can then be calculated by integrating  $\mathbf{m}(\mathbf{r},t)$ over space, $\mathbf{r}$. 
This method of determining the magnetic moment represents the gold-standard within TD-DFT, in the sense that it is completely free from approximation for a given exchange-correlation functional.
In Scheme (ii) the magnitudes of the local magnetic moments can be calculated from the responses of the material to probe pulses (x-ray beams).
Such a response is given by a dielectric response tensor, $\epsilon$, which is the change in density of the material upon perturbation by the probe pulse.
Scheme (ii) follows a pump-probe experimental technique for obtaining the moments, but we compute rather than measure the transient response for various pump-probe delays.
Our method of calculating the transient response function, $\epsilon$, of a pumped system at the $M$ and $L$ resonant absorption edges within TD-DFT has already been studied extensively and demonstrated to be in very good agreement with experiment \cite{willems_optical_2020,dewhurst2020,sharma_computational_2022}. 
For magnetic linear dichroism (MLD) the relation between the response and the magnetisation is \cite{kunes_understanding_2004}, 
\begin{equation}
\mathbf{M}^2 \propto \Im (\epsilon_\parallel - \epsilon_\perp),
\label{eqn:MLD}
\end{equation}
where $\parallel$ and $\perp$ are defined relative to the direction of the magnetic moment $\mathbf{M}$ (for example $\epsilon_\parallel$ is the dielectric response parallel to the direction of $\mathbf{M}$). For MCD a different relationship exists, see Ref.~\onlinecite{sharma_computational_2022,
Stohr2006}.

\emph{Computational details}:
All calculations were performed using the highly accurate full potential linearized augmented-plane-wave method\cite{singh}, as implemented in the ELK\cite{elk,dewhurst2016} code. 
A face centred cubic unit cell with lattice parameter of $3.21\si{\angstrom}$ was used for AFM FePd and AFM FePt, while for FM FePt a lattice parameter of $3.53\si{\angstrom}$ was used. 
The magnitudes of the groundstate Fe moments are identical in AFM FePd and FM FePt, with a value of 2.77~$\mu_{\rm B}$, where in the case of FePt the Pt atom acquires a small induced moment of 0.35~$\mu_{\rm B}$.

\begin{figure}[t!]
\includegraphics[width=0.9\columnwidth, clip]{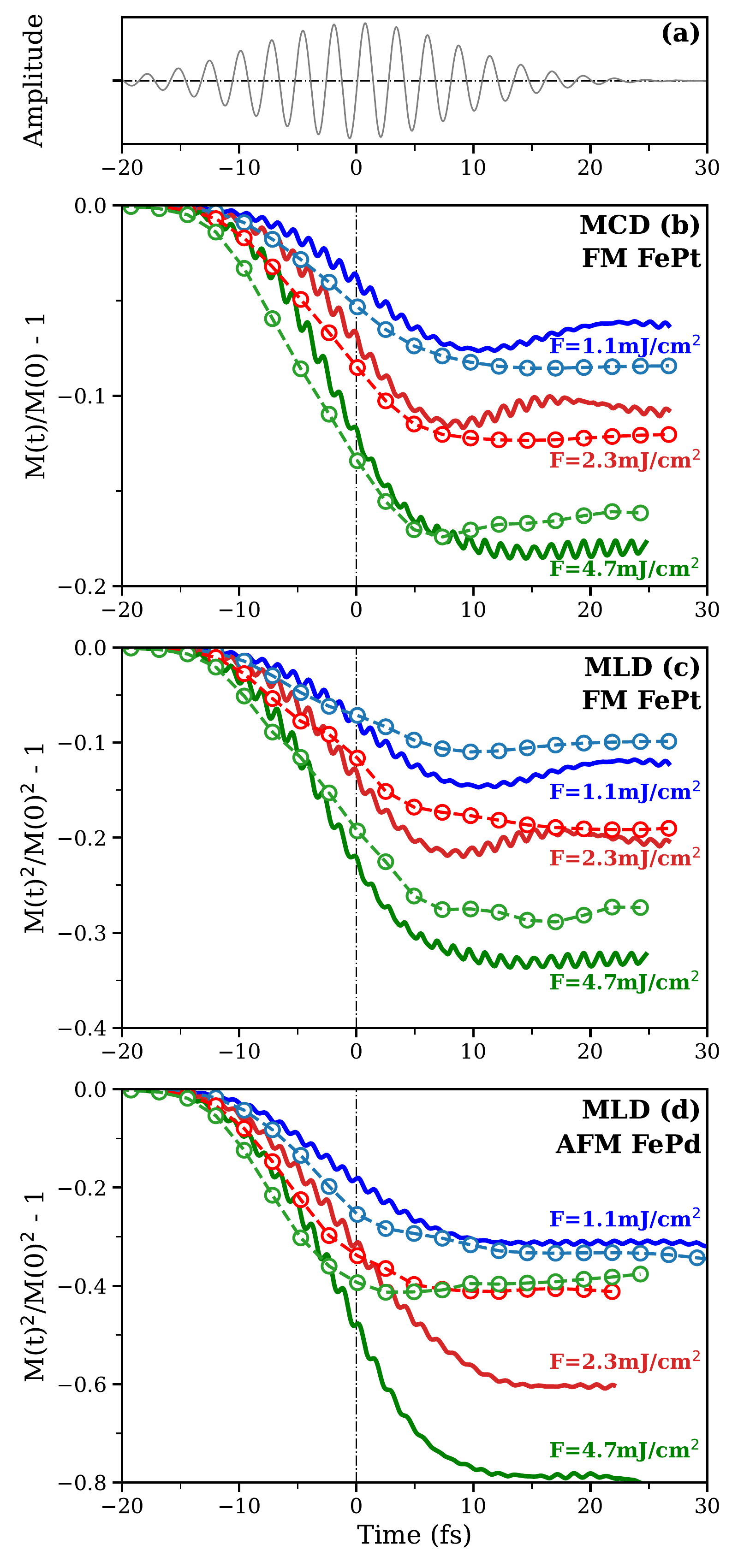}
\caption{{\it Transient moments obtained (i) directly via the spin density and (ii) via the $L_3$ edge response functions in laser pumped FePt and FePd}. The pump laser electric field is shown in panel (a), with in all results the frequency (1.55~eV) and duration (24~fs) held fixed and fluence was modified through changing the amplitude.
The transient normalized moment is shown for laser pumped ferromagnetic FePt (b), with the transient normalized moment squared shown for ferromagnetic FePt (c), and antiferromagnetic FePd (d). In each case the solid lines denote the moment obtained from spin density and the open circles that obtained via the $L_3$ edge response. Evidently while magnetic circular dichroism (MCD) and linear dichroism (MLD) capture well the transient moment in ferromagnetic FePt, in antiferromagnetic FePd the failure of MLD becomes significant at a fluence of 2.3~mJ/cm$^2$.
}\label{fig:FePt_FePd_MLD}
\end{figure}


The Brillouin zone was sampled with a $10\times 10\times 8$ $k$-point mesh for both materials.
To calculate the spin dynamics we have used the {\it ab-initio} state-of-the-art fully non-collinear spin-dependent version \cite{krieger2015,dewhurst2016} of time dependent density functional theory (TDDFT)\cite{RG1984} with the adiabatic local spin density approximation for the exchange-correlation potential. Spin-orbit coupling was included in all calculations.
For time propagation the algorithm detailed in Ref.~\onlinecite{dewhurst2016} was used with a time-step of $2.42$~attoseconds. We evaluate the response functions at the L$_3$ peak\cite{kunes_understanding_2004}, for which we employ the bootstrap kernel\cite{sharma2011} to treat excitonic effects, with a smearing of 0.9~eV, a value derived from the $GW$-calculated average width of the semi-core $p_{3/2}$ and $p_{1/2}$ $L$ peaks. By use of a scissor operator the $2p$ Kohn-Sham states of the DFT calculations were rigidly shifted in energy according to the outcome of the $GW$ calculations, and these energy-corrected states were used in calculations of the response functions. The $GW$ calculations were performed at a temperature of 500~K using the all-electron, spin-polarized $GW$ method\cite{gw} as implemented in the Elk code\cite{elk}. A Matsubara cutoff of 12~Ha was used. The spectral function on the real axis was constructed using a Pade approximation \cite{vidberg_solving_1977}. We find that, in contrast to spectra obtained at the $M$-edge\cite{dewhurst2020}, local field effects are not important for the $L$-edge spectra that we study here.

\begin{figure}[t!]{
\includegraphics[width=0.85\columnwidth, clip]{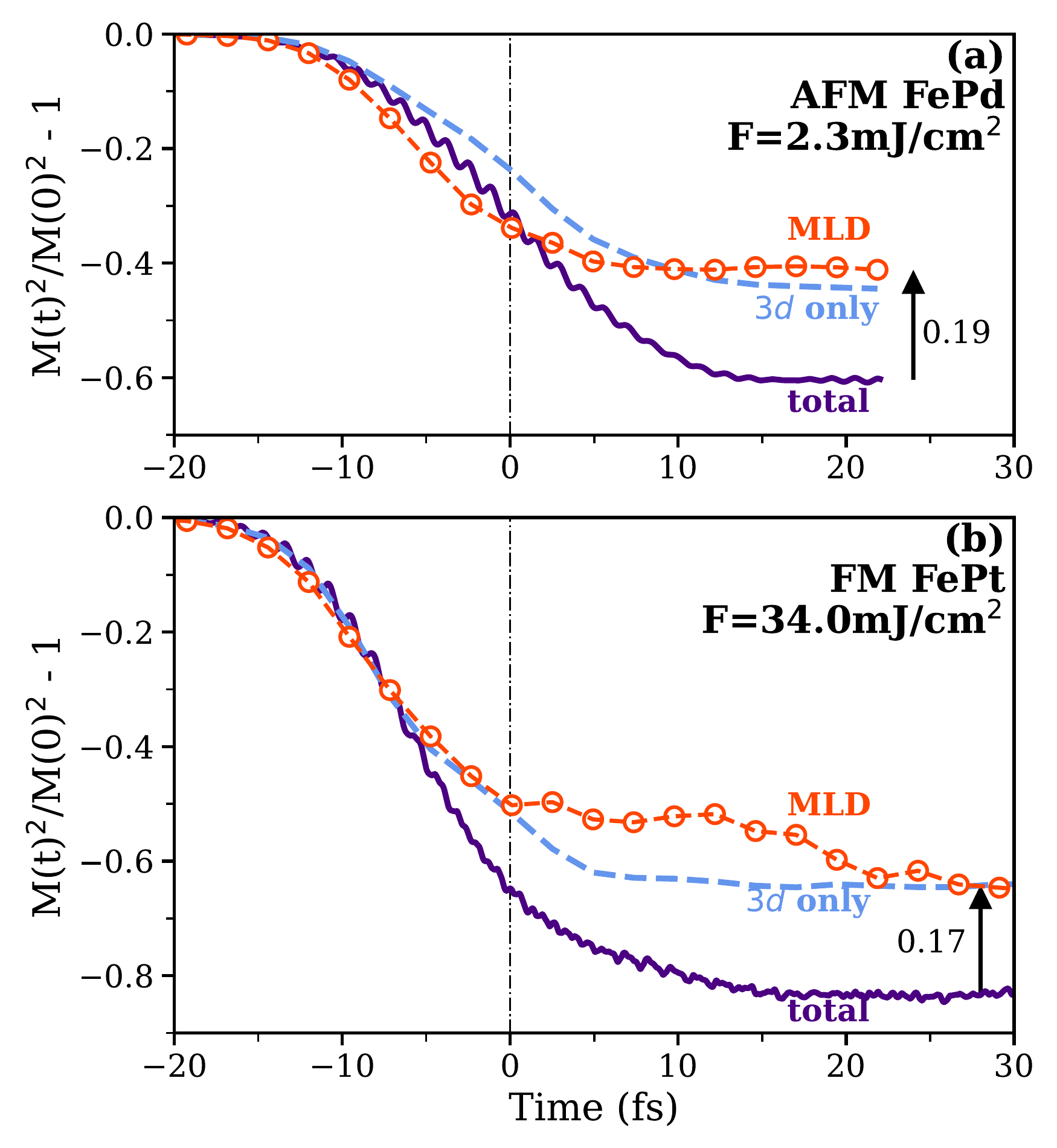}
}
\caption{{\it Response function methods fail for the total transient moment, but capture well the $d$-band moment}.
Shown is the square of the transient transient magnetic moment (normalized as indicated on the axis) for the total Fe moment (purple, solid line), the $d$-band moment (light blue, dashed line), and the MLD at the L$_3$ edge derived moment (orange, rings) in laser pumped antiferromagnetic FePd (a) (matching the red data in Fig.~\ref{fig:FePt_FePd_MLD} (c)), and ferromagnetic FePt (b). Note the dramatically increased fluence (34.0~mJ/cm$^2$ as opposed to 2.3~mJ/cm$^2$) but also the excellent agreement of MLD with the $d$-band moment even when it grossly fails to capture the total transient moment. The pulse frequency and duration employed is in each case that presented in Fig.~\ref{fig:FePt_FePd_MLD} (a).
}\label{fig:FePt_FePd_d-char}
\end{figure}

\emph{Results:}
Firstly, to give important context to the  assessment of the error of MLD for FePt and FePd, we compare the accuracy of MLD to the that of MCD for a range of pump pulse fluences. In each case the central frequency (always 1.55~eV) and the duration (24~fs) are held fixed (see Fig.~\ref{fig:FePt_FePd_MLD}(a)).
This comparison is possible only for materials with FM ordering, such as FePt, since MCD, proportional to $\v M$, is identically zero for AFM materials. 
Our calculations of the time-evolution of  $M$ and $M^2$ for FM FePt during and after the pump laser pulse are shown in Fig.~\ref{fig:FePt_FePd_MLD}(b) and (c) respectively. 
For each pulse fluence we present data obtained both from Scheme (i), evaluation via the spin operator, as well as from Scheme (ii), calculation via the $L_3$ edge response functions.
Whilst our two distinct schemes do not produce identical results, the similarity is sufficient to conclude a meaningful evaluation of the transient magnetisation could be made from the dichroism of the response (MLD or MCD). For both MLD and MCD we also find an increasing deviation of the two schemes as pump fluence is increased further beyond the displayed values, consistent with the finding observed in the case of MCD applied to elemental ferromagnets
\cite{sharma_computational_2022}.
However, in stark contrast to this previous finding \cite{sharma_computational_2022} for elemental ferromagnets we find here the moment is \textit{over}estimated by the dichroism, not underestimated.

Having established the accuracy of MLD as a probe in FM FePt for a range of pump fluences, we now turn to the important case of AFM FePd. Our calculations of $M^2$ for the Fe atom in AFM FePd are shown in Fig.~\ref{fig:FePt_FePd_MLD} (d). 
We note that for 4.7~mJ/cm$^2$ incident fluence AFM FePd absorbs approximately 15\% more energy while 40\% more charge is excited.
In striking contrast to the MLD results for FM FePt (Fig.~\ref{fig:FePt_FePd_MLD} (c)), the failure of the MLD to accurately track the magnetisation in AFM FePd is dramatic: only at the lowest incident fluence (1.1~mJ/cm$^2$) is the difference between the two methods for obtaining $M^2$ tolerable.
For the pulses with 2.3~mJ/cm$^2$ and 4.7~mJ/cm$^2$ incident fluence direct evaluation of $M^2$ reveals far greater demagnetisation than indicated by the MLD, with for the 4.7~mJ/cm$^2$ case the MLD normalized $M^2$ amounting to only 50\% of the directly calculated value, equivalent to an overestimation of the magnitude of the Fe moments by 70\%. The message of these results is thus clear: MLD is, somewhat unexpectedly, profoundly less accurate for AFM FePd than for FM FePt.

We find that, as is the case for MCD\cite{sharma_computational_2022}, the failure of MLD occurs as transient moment is not only reduced by the pump laser, but also {\it delocalized}. This is shown in Fig.~\ref{fig:FePt_FePd_d-char}(a) in which can be seen that the $M^2$ of $3d$ character significantly differs from the total transient $M^2$, indicating a strong delocalization. This invalidates the central assumption underpinning the response based moment calculation, that of the $d$-band state counting\cite{Stohr2006}, where this finding is reinforced by the fact that the transient $d$-band $M^2$ agrees very well with the transient $M^2$ obtained from the MLD calculation.

The remarkable scale of the difference in error between FM FePt and AFM FePd can be made clear by considering the incident power density required to reproduce, in FePt, the error found in FePd at a fluence of 2.3~mJ/cm$^2$. As can be seen in Fig.~\ref{fig:FePt_FePd_d-char}(b) this requires increasing the fluence by more than an order of magnitude to 34~mJ/cm$^2$ (with the absorbed energy $\sim 8$ times larger).

\begin{figure}[t!]{
\includegraphics[width=0.9\columnwidth, clip]{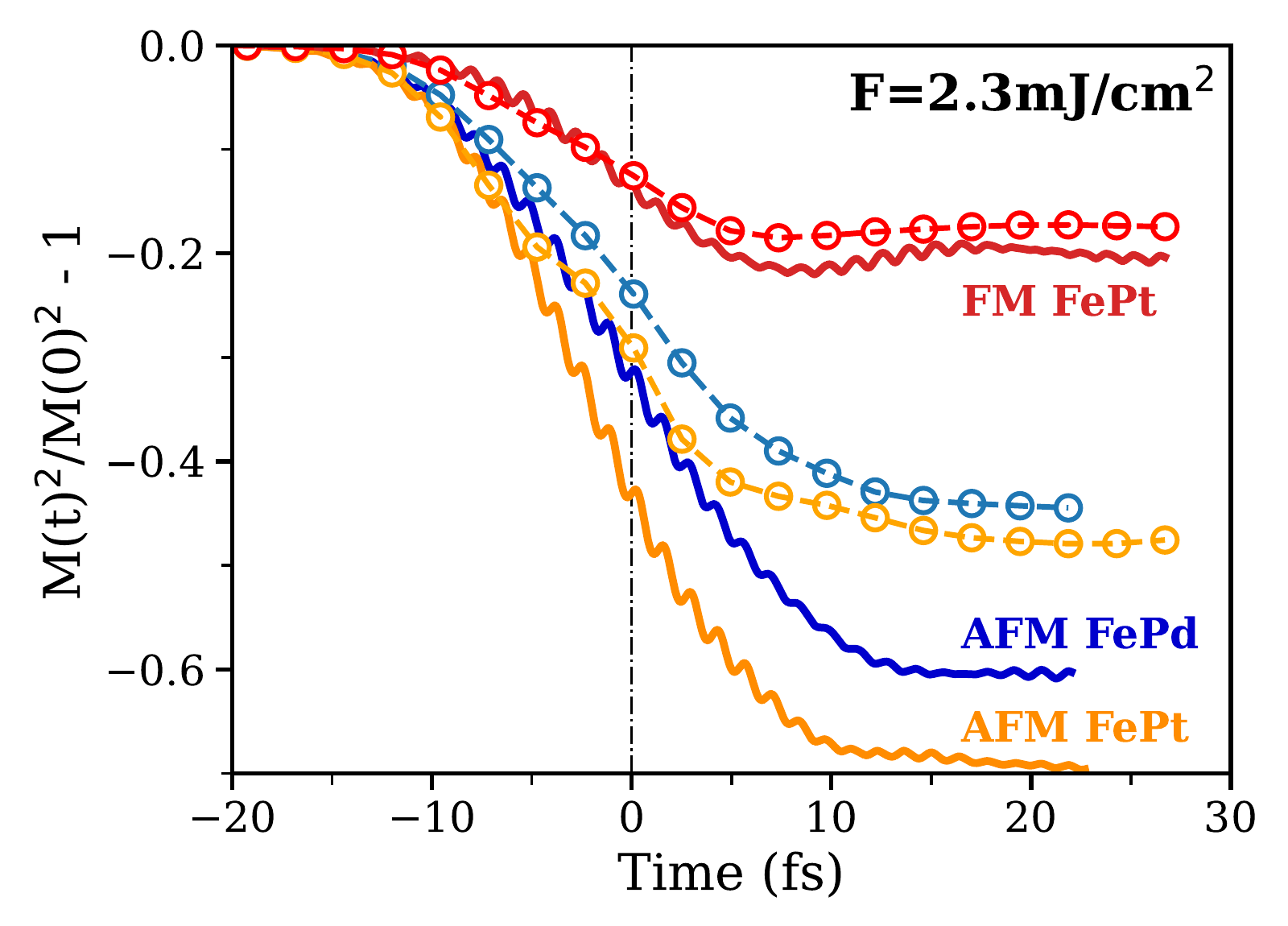}
}
\caption{{\it Primary role of magnetic structure in the accuracy of MLD as a probe of demagnetization}. Shown is the square of the transient moment (normalized as indicated on the axis), calculated both from the spin density and from the MLD response of the L$_3$ edge in laser pumped ferromagnetic FePt (red), antiferromagnetic FePd (blue), and antiferromagnetic FePt (yellow). Evidently, enforcing an antiferromagnetic order in FePt results in a failure of the MLD probe very similar to that found in FePd. The pump pulse is that presented in Fig.~\ref{fig:FePt_FePd_MLD}(a) with the fluence set to  2.3~mJ/cm$^2$.
}\label{fig:FePt-AFM}
\end{figure}

We now wish to understand the origin of this remarkable difference on an even more fundamental level. First, we seek to thoroughly define which of the distinctions between FM FePt and AFM FePd might be important. Pt and Pd are isoelectronic elements, and FePt shares its crystal structure with FePd, so the clearest point of distinction is the magnetic ordering.
Nevertheless, we wish to rigorously exclude the possibility that the difference between Pt and Pd could be an important factor, so we now show in Fig.~\ref{fig:FePt-AFM} a calculation of the magnetization dynamics of FePt when it is artificially constrained to an AFM coupled state.
This conclusively demonstrates that the higher error in AFM FePd is a direct result of the different magnetic coupling.

\begin{figure}[t!]{
\includegraphics[width=0.99\columnwidth, clip]{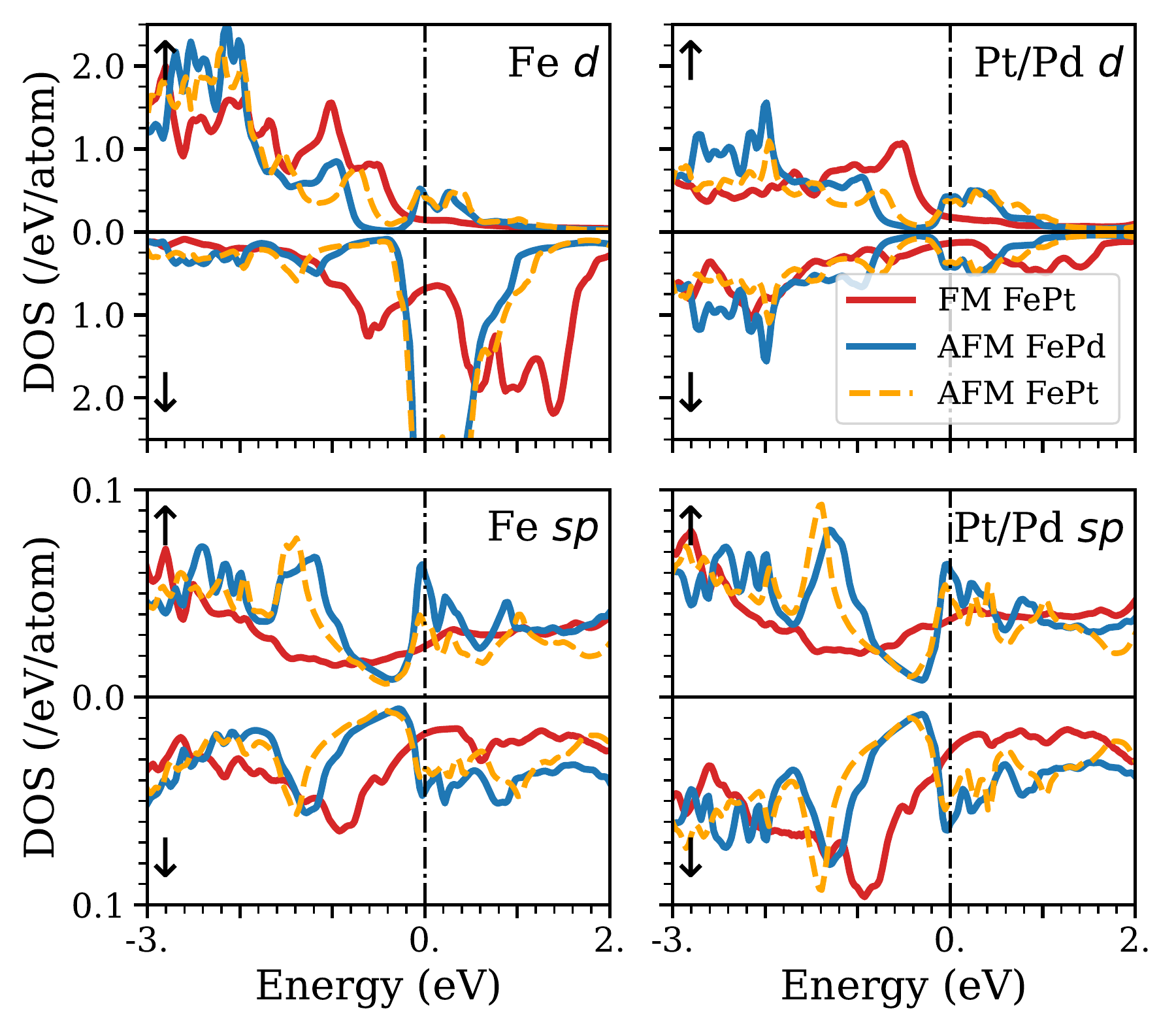}
}
\caption{{\it Influence of magnetic ordering on $sp$-$d$ hybridization}.
The equilibrium density of states (DOS) for ferromagnetic FePt (blue), antiferromagnetic FePt (light blue, dashed) and antiferromagnetic FePd (orange), with in each case the Fermi energy set to zero. The species and angular momentum resolved DOS, as indicated by the labels of the separate panels, reveal an increase in $sp$-states close to the Fermi energy that, along with the peak in minority $d$-states allows increased pump laser excitation of minority electrons away from $d$ states. As described in the text, this results in significant underestimation of demagnetization as recorded in a MLD pump-probe setup.
Note that for the case of antiferromagnetic FePd and FePt the available spin up ($\uparrow$) and down ($\downarrow$) states are reversed on the second Fe atom (not shown).
}\label{fig:PDOS}
\end{figure}

Next, we must ask why the error is higher, or, equivalently, why a greater amount of charge (and magnetisation) is excited to outside the $d$-states if there is AFM coupling.
AFM materials have previously been shown to demagnetise faster and to a greater extent, because AFM ordering permits a demagnetisation mechanism whereby charge can be driven rapidly from occupied majority states on one atom to the unoccupied minority states on the other (an OISTR \cite{thielemann-kuhn_ultrafast_2017,dewhurst_laser-induced_2018}).
It would thus be simplest to assume that the greater error is simply a result of this greater, faster, demagnetisation.
In actual fact, we find that this would be a false assumption: the size of the error is not simply proportional to the size of the magnetisation quenching, as can be seen by comparison of Fig.~\ref{fig:FePt_FePd_d-char}(a) and (b).
Rather, we can clarify the reason for the greater amount of charge excitation to non-$d$-character states, and therefore understand the error level, through a consideration of the energy-dependent density of states, which is shown in Fig.~\ref{fig:PDOS} for FM FePt, AFM FePd, and AFM FePt, where contributions to the total density of states from states with each atom and angular momentum character are shown separately.

The clearest difference between the FM and AFM materials is the presence of the AFM pseudogap, with associated magnetic bonding and anti-bonding exchange split $d$-bands, at -2.2~eV and 0.5~eV respectively. This results in a very high peak in minority $d$ states on the Fe atom in the energy domain at the Fermi energy and up to 1~eV above it. Crucially, driven by the marked reduction of states in the pseudogap, a significant increase in $sp$-states is also found in the same energy window (more than double that found for the case of FM order), and it is the concomitant increase in allowed $d\to sp$ transitions that underpins the increased delocalization seen in AFM FePd when a pump pulse is applied. Having its origin in the pseudogap associated with sub-lattice magnetic ordering (see e.g. Ref.~\onlinecite{PhysRevB.76.054444} or Ref.~\onlinecite{Kubler2000} for a discussion of the importance of the pseudogap in magnetic order) the dramatically reduced threshold for failure of MLD in FePd will, likely, be a general feature of transition metal antiferromagnets.

\emph{Conclusion:}
Our investigation of FePt (a ferromagnet) and FePd (an antiferromagnet) has revealed profoundly different failure thresholds exist for response function based methods of determining the transient moment, with massive ($>$50\%) MLD errors being generated in FePd at fluences (4.7~mJ/cm$^2$) for which MLD and MCD capture very well the transient moment in FePt.
This failure results from the fact that pump laser pulses not only reduce but also {\it delocalize} the angular momentum of magnetic order, and in FePd the AFM pseudogap, that results in significantly increased $d$-$sp$ hybridization, aids this delocalization. As response function based methods rely on a $d$-band state counting principle, the excitation of significant moment to states of non-$d$-character guarantees their failure; in FePd the greater tendency to excite out of the $d$-band thus results in failure to record the total transient moment at much lower pulse intensity. A corollary of this, however, is that such methods will always capture the $d$-band moment, and our calculations confirm this: the MLD and MCD derived moments reproduce excellently the underlying transient $d$-band moments at all pulse intensities. Importantly, therefore, a valid protocol exists for collaboration between theory and experiment: comparison via the transient $d$-band moment. In summary, our work highlights the critical role that electronic state hybridization has in affecting the accuracy of MLD and MCD in tracking transient moments in ultrafast pump laser experiments, suggesting that a cautious approach is needed in the interpretation of experimental spectra as unveiling total transient moments.

\emph{Acknowledgements}: E. I. Harris-Lee, J. K. Dewhurst and S. Sharma would like to thank the DFG for funding through project-ID 328545488 TRR227 (project A04).
S. Shallcross and S. Sharma would like to thank Leibniz Professorin Program (SAW P118/2021) for funding. 
The authors acknowledge the North-German Supercomputing Alliance (HLRN) for providing HPC resources that have contributed to the research results reported in this paper.

\bibliography{Ledge,Tr-MLD-paper}

\end{document}